\documentstyle[aps,epsf,prl,multicol]{revtex}
\draft

\begin{document}

\title{Singularity in the boundary resistance between superfluid $^4$He and a solid surface} 

\author{Kerry Kuehn$^*$, Sarabjit Mehta$^*$, Haiying Fu$^*$, Edgar Genio$^*$, Daniel Murphy$^*$, \\Fengchuan Liu$^\dagger$, Yuanming Liu$^\dagger$, and Guenter Ahlers$^*$}

\address{$^*$Department of Physics and iQUEST,\\ University of
California, Santa Barbara, CA  93106, USA\\ $^\dagger$Jet Propulsion Laboratory, California Institute of Technology, Pasadena, CA 91109}

\date{\today}\maketitle

\begin{abstract}
We report new measurements in four cells of the thermal boundary resistance $R$ between copper and $^4$He below but near the superfluid-transition temperature $T_\lambda$. For $10^{-7} \leq t \equiv 1 - T/T_\lambda \leq 10^{-4}$ fits of $R = R_0 t^{x_b} + B_0$ to the data yielded $x_b \simeq 0.18$, whereas  a fit to theoretical values based on the renormalization-group theory yielded $x_b = 0.23$. Alternatively, a good fit of the theory to the data could be obtained if the {\it amplitude} of the prediction was reduced by a factor close to two. The results raise the question whether the boundary conditions used in the theory should be modified.

\end{abstract}

\begin{multicols}{2}

An important, as yet unresolved, issue in condensed-matter physics is the nature of the boundary conditions which are appropriate to describe the interface between a liquid and a solid. Experimentally it is difficult to investigate this problem because the boundary layer usually is of atomic dimensions and contributes negligibly to macroscopic properties. This situation is changed near a critical point, where the thickness of a boundary layer is of the same order as the correlation length $\xi$; the correlation length diverges at the critical temperature. A system particularly suitable for the investigation of the boundary layer is the superfluid transition of $^4$He.

\begin{figure}
\epsfxsize=2.25in
\centerline{\epsffile{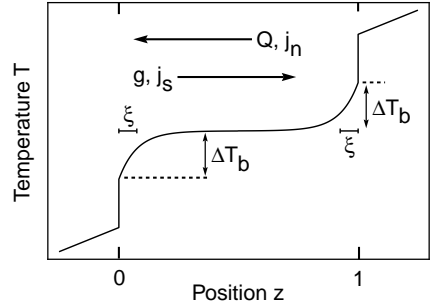}}
\vskip 0.2in
\caption{Schematic diagram (not to scale) of the temperature variation in the cell. The discontinuity at $z = 0$ ($z = 1$) is due to the classical Kapitza resistance $R_K$ at the top (bottom) of the cell, and $\Delta T_b$ (which develops over a length scale $\xi$) is due to $R_b$. The directions of the currents $Q, j_n, j_s$, and of gravity $g$ are indicated.}
\label{fig:schematic}
\end{figure}

In bulk superfluid $^4$He heat transport occurs through the counterflow of the normalfluid ($j_n$) and superfluid ($j_s$) currents and does not produce a temperature gradient \cite{La41}. However, Landau predicted \cite{La41} that a heat current $Q$ orthogonal to a solid wall should produce a boundary layer in the fluid adjacent to the wall with a temperature difference $\Delta T_b$ across it. The reason for this is the suppression of the order parameter, and thus of $j_s$ and $j_n$, near the wall. The phenomenon is illustrated schematically in Fig.~\ref{fig:schematic}. Within this boundary layer, suppression of the counterflow implies that some of the heat must be carried by diffusive processes.  Consequently, a thermal gradient is developed within it. However, the thermal resistance 
$
R_b = \Delta T_b / Q
$
is unobservably small deep in the superfluid phase where the boundary-layer thickness is of atomic dimensions. Only very close to the superfluid transition temperature $T_\lambda$ where $t = 1 - T/T_\lambda$
becomes small and $\xi = \xi_0 t^{-\nu}$ becomes large does $R_b$ become measurable in high-precision experiments. Thus this phenomenon was found experimentally only relatively recently \cite{DAS87,DA91,ZTM90}. Roughly speaking one can assume that the temperature gradient in the boundary layer decays exponentially from $-Q/\lambda$ at the wall (where there is presumed to be no counterflow at all) to zero deep in the superfluid (where all the heat is carried by counterflow) with a characteristic length equal to $\xi$, as illustrated in Fig.~\ref{fig:schematic}. Here  $\lambda \simeq \lambda_0 t^{-x_\lambda}$,  with an effective exponent $x_\lambda \simeq 0.43$, is the diffusive thermal conductivity of the fluid \cite{Ah68}. This leads to the estimate 
\begin{equation}
R_b \simeq \xi/\lambda \propto t^{-x_b}
\label{eq:R_b}
\end{equation}
where $x_b = \nu - x_\lambda \simeq 0.24$ \cite{effective}. A renormalization-group-theoretical calculation of $R_b$ carried out by Frank and Dohm (FD) \cite{FD89,FD91} agrees with this qualitative picture and is expected to give the behavior of $R_b(t)$ quantitatively.   

\narrowtext

The singular boundary resistance is particularly interesting because it is a direct consequence of the boundary conditions (BC) which are relevant at the solid-liquid interface. For instance, theoretically popular (but physically unrealistic) BC in the study of finite-size effects near critical points are periodic ones (PBC) \cite{Ba83}. For PBC, there is {\it no} contribution in the fluid to the boundary resistance at the solid-liquid interface because the order parameter is not suppressed. More realistic BC are of the Dirichlet type (DBC). In that case it is assumed that the order parameter {\it vanishes} at the solid wall. This is the case considered by FD and illustrated in Fig.~\ref{fig:schematic}. In principle, one can imagine a continuous family of BC \cite{FGDL88}, and an important question which goes well beyond the problem of critical phenomena is which BC are relevant to real physical systems. 

We find that excellent fits to new data for four separate cells (labeled I to IV) can be obtained with the  powerlaw Eq.~\ref{eq:R_b}, but  that the resulting values of $x_b$ are close to 0.18 and thus significantly smaller than the theoretically-expected value $x_b^{th} \simeq 0.24$. Alternatively, a reasonable fit of the function predicted by the theory \cite{FD91} to the data can be obtained if this function is multiplied by a factor which is significantly less than one and, for three of the cells, close to 0.53. Although a detailed theoretical explanation of this problem may be required, we believe that this discrepancy may be indicative of a need to consider modifications of the DBC which have been assumed in the theory not only of $R_b$, but also of finite-size effects on the specific heat \cite{Do}.  

The apparatus was similar to one described previously \cite{DA91,SA84a}. We used four different sample cells. Cell I was described elsewhere \cite{BAKF00,FBKA98}. It had a diameter of 1.27 cm. The 0.12 mm thick stainless steel sidewall was brazed to a top copper piece, leaving no gap between it and the copper \cite{gap}. The copper surface at the sample top was finely machined and had a measured root-mean-square (rms) roughness of 1 $\mu$m. The fill capillary entered the cell bottom.  Two sideplanes (SPs) similar to those first used by Duncan et al. \cite{DAS87,DA91} were used to measure the helium temperature. They pressed around the entire circumference against the outside surface of the sidewall. Thermometers on the SPs could sense the helium temperature, unencumbered by any temperature drop across a solid-liquid interface. Cells II and III had a similar construction. However, the cell {\it bottom} was sealed to the {\it bottom} copper piece with epoxy (brazing) for cell II (III), and the fill lines entered the cells from the top. The bottom surfaces were polished copper and the rms surface roughness was 0.01 (0.05) $\mu$m for cell II (III).  
Cell IV was similar to one used in the microgravity project DYNAMX \cite{DE01}. It had copper top and bottom cylinders of diameter 2.28 cm which were epoxied into a stainless steel wall of thickness 0.075 mm. The two copper surfaces were polished and had a rms roughness of about 0.02 $\mu$m. The wall was {\it penetrated} by two copper SPs which made contact all around the circumference with the helium. The thickness of each SP was 0.075 mm. The fill capillary entered the cell top. 

The temperatures of the top ($T_t$), bottom ($T_b$), and the two sideplanes ($T_{st}$ and $T_{sb}$) were measured with nano-K resolution \cite{FBKLMSA98}. In each case the fill capillary came from an overflow volume containing the liquid-vapor interface on an isothermal shield which was kept at a temperature near 2.2 K and steady to $\pm 1 \mu$K \cite{exception}. The surface {\it not} penetrated by the fill capillary was used for the measurements. The sample was $^4$He with less than 1 ppb $^3$He. 

The calculation of Frank and Dohm \cite{FD91} gives the difference $\delta R_b^{th} = R_b^{th}(t) - R_b^{th}(t_0)$ and thus the values of $R_b^{th}$ are determined only to within an additive constant. The theoretical result has the form
\begin{equation}
\delta R_b^{th} = \int\limits_{2t}^{2t_0}{dt' \over t'}{\xi \over {2 \lambda}} K(f,w,u, \gamma) \ \ .
\label{eq:Rb_th}
\end{equation}
Here $\lambda$ and $\xi$ are the thermal conductivity and correlation length above $T_\lambda$, and $K$ is a function of the static and dynamic coupling constants $\gamma$, $u$, $f$, and $w$ of Model F \cite{HHS76} for the dynamics of the superfluid transition. The coupling constants depend on $t$ and are known from independent fits of the theory \cite{Do91} to the thermal conductivity \cite{TA85} and heat capacity \cite{TA85,Ah71} near $T_\lambda$. Using their tabulated values \cite{DM91}, we evaluated  $\delta R_b^{th}$ numerically from the equations given in Ref.~\cite{FD91} and $t_0 = 0.05$. 

\begin{figure}
\epsfxsize=3.in
\centerline{\epsffile{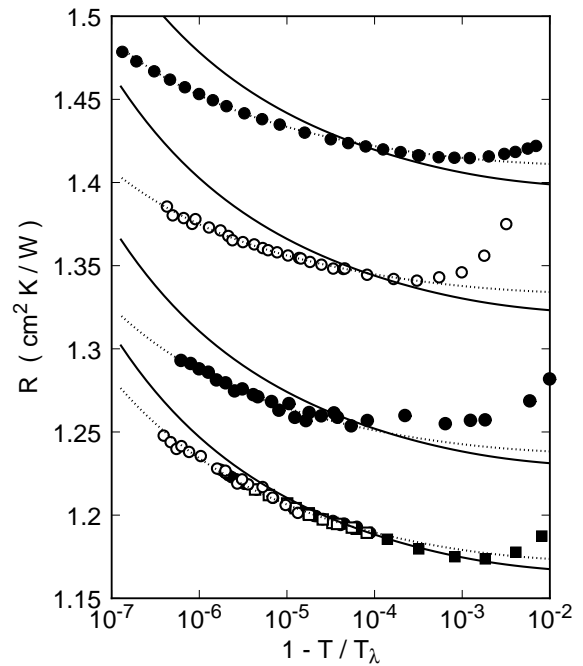}}
\vskip 0.2in
\caption{The resistance $R$ of (from bottom to top) cell I (moved up by 0.1 cm$^2$ K/W), II, III, and IV (moved down by 0.2 cm$^2$ K/W). For cell I the open circles, solid circles, open squares, and solid squares are for data taken at 2, 20, 30, and 40 $\mu$W/cm$^2$ respectively. The solid lines are the theoretical result, shifted vertically so as to agree with the data near $t = 10^{-4}$. The dotted lines are the theory multiplied by factors of 0.76, 0.59, 0.51, and 0.52 for cell I to IV respectively and shifted vertically. }
\label{fig:linear}
\end{figure}

In Fig.~\ref{fig:linear} the data for $R = R_b + R_B$ are shown for all four cells on a linear vertical scale. For cell I there are data for several power densities $Q$ which show that, within the experimental resolution, the results do not depend on $Q$ \cite{gap}.  There is a large sample-dependent background contribution $R_B = R_K + R_{Cu}$  due to the usual Kapitza resistance $R_K$ and the series resistance $R_{Cu}$ in the copper parts of the cell (see Fig.~\ref{fig:schematic}). The theoretical prediction \cite{FD91} of $R_b$, after vertical shifts which cause agreement with the data near $t = 10^{-4}$, is given by the solid lines in the figure. Obviously the agreement is not good.
 
\begin{figure}
\epsfxsize=2.75in
\centerline{\epsffile{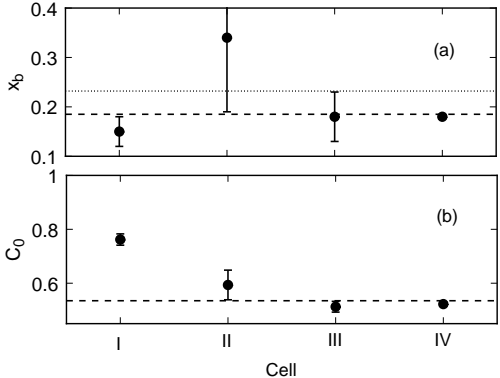}}
\vskip 0.2in
\caption{(a): Values of $x_b$ derived from fitting Eq.~\ref{eq:powerlaw} to the data for cells I to IV. The dotted line is the theoretical value, and the dashed line is an average value based on the experiments. (b): Values of $C_0$ derived from fitting Eq.~\ref{eq:FDfit} to the data for cells I to IV. The theoretical value corresponds to the top of the graph at $C_0 = 1$. The dashed line is an average value for cells II to IV. }
\label{fig:paras}
\end{figure}
 
Equation~\ref{eq:R_b} suggests that it should be possible to fit the data near $T_\lambda$ with a powerlaw. Thus, as an alternative interpretation we considered the data for very small $t$ where we can represent $R_B$ by a constant. We fitted the function
\begin{equation}
R = R_0 t ^{-x_b} + R_B
\label{eq:powerlaw}
\end{equation}
to the data with $t < 10^{-4}$, least-squares adjusting $R_0, x_b,$ and $R_B$. For $t \equiv 1 - \bar T / T_\lambda$ we used the zero-current value of $T_\lambda$ at the vertical position of the cell surface in question \cite{Ah91} and $\bar T \equiv T_{He} + \Delta T_b/2$. We obtained the values of $x_b$ shown in Fig.~\ref{fig:paras}a. The error bars are 95\% confidence limits derived from the random errors and do not include possible systemstic errors. An equivalent fit to the theoretical data yielded $x_b^{th} = 0.232$ which is shown as a dotted horizontal line in the figure. We see that the uncertainties of $x_b$ are largest for cells II and III, and that the data for those cells are consistent with the theoretical value.  
However, the more precise data for cells I and IV are clearly inconsistent with the theory. All four experimental results are consistent with $x_b \simeq 0.185$ which is shown as a dashed horizontal line in the figure. For comparison, we mention that a similar fit to the data from cell J of Ref.~\cite{DA91} yielded $x_b = 0.14 \pm 0.54$ \cite{whynot}. One sees that the uncertainty in these data is large enough for them to be consistent with the theory.   
  
\begin{figure}
\epsfxsize=2.75in
\centerline{\epsffile{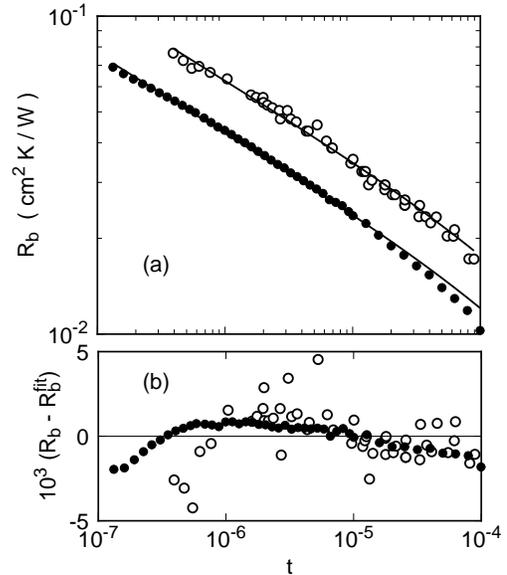}}
\vskip 0.2in
\caption{Upper panel: The contribution $R_b = R - B_1$ with $B_1$ determined from a fit of Eq.~\ref{eq:FDfit} to the data. Open (solid) circles are for cell I (IV) with $B_1 = 1.072 ~ (1.610)$. The solid lines are the fits. (b): Deviations of the data from the fits.}
\label{fig:FDfit}
\end{figure}
   
In order to learn more about the difference between theory and experiment, we considered the possibility that the theory might give the $t-$dependence of $R_b$ correctly, but that the amplitude is in error by a factor of order unity. Thus we fitted the function 
\begin{equation}
R = C_0 \delta R_b^{th} + B_1
\label{eq:FDfit}
\end{equation} 
to the data over the range $t \leq 10^{-4}$. Here $B_1$ represents the sum of the contributions from the experimental background  $R_B$ and from the unknown contribution $R_b^{th}(t_0)$ in the theory (see Eq.~\ref{eq:Rb_th}). Both $C_0$ and $B_1$ were least-squares adjusted. We found that the data could be fitted reasonably well, but with $C_0 < 1$. We obtained the values of $C_0$ shown in Fig.~\ref{fig:paras}b (the error bars again are 95\% confidence limits). The fit to the data is quite good, as shown by the dotted lines in Fig.~\ref{fig:linear}. The data from cell J of Ref.~\cite{DA91} yielded $C_0 = 0.88 \pm 0.26$ from a similar fit \cite{whynot}. As can be seen, the error bars based on these older data easily overlap the theoretical result $C_0 = 1$. 

For all four cells, $C_0$ is significantly less than unity. The results for cells II to IV are consistent with $C_0 \simeq 0.535$, which is shown by the dashed line in Fig.~\ref{fig:paras}b. This result is smaller than expected from estimates of the theoretical uncertainty of the amplitude of $R_b^{th}$. We do not know why the result for cell I is somewhat higher, but note that cell I was the only one which had a relatively rough and unpolished surface. To illustrate the quality of the fit to the data, we show  
$R_b = R - B_1$ and $C_0 \delta R_b^{th}$ for cells I and IV (which have the highest precision) in Fig.~\ref{fig:FDfit}a, and the deviations of the data from $C_0 \delta R_b^{th}$ in Fig.~\ref{fig:FDfit}b. We see that the fit to the data, although quite good,  is not perfect particularly for cell IV which yielded the most precise data. 

We note that the theoretical result was obtained from a first-order perturbation calculation. It would be natural to look here for the difference between theory and experiment. However, the value of $K$ in Eq.~\ref{eq:Rb_th} is nearly independent of $t$, leaving the exponent of $R_b$ essentially unaltered and nearly equal to that of $\xi/\lambda$. It seems unlikely that this situation would change at higher order. The {\it value} of $K$ might change by a factor of order one and thus affect $C_0$ if a higher-order calculation were performed; but a change by a factor as small as 0.53 seems unlikely. This leaves us to suspect the choice of the DBC as the cause of the difference between the theory and the experiment. It would be very interesting to investigate theoretically whether the BC can be modified so as to cause the calculation to match the experimental results more closely.

This work was supported by NASA Grants NAG8-1429 and NAG8-1757.

\end{multicols}


\begin{references}
\bibitem{exception}For cell IV the data well below $T_\lambda$ were taken with the overflow volume slightly below $T_\lambda$ to reduce the heat leak to the cell to an acceptable level.
\bibitem{whynot} We did not include the cell K data from Ref.~\cite{DA91}. They agree well with the cell J data (see Fig.~7 of Ref.~\cite{DA91}), but cell K had a sidewall gap (see Ref.~\cite{gap}) which makes the effective area for the determination of $R$ uncertain.
\end{references}
\end{document}